\documentclass[12pt]{iopart}

\usepackage{graphicx}
\usepackage{xcolor}
\usepackage{tikz}
\usepackage{bm}
\usepackage{algorithmic}
\usepackage{algorithm}
\usepackage{hyperref}       
\usepackage{comment}
\usepackage{booktabs}
\usepackage{multirow}
\newcommand{\diff}{\mathrm{d}}

\usepackage{CJKutf8}

\begin{document}
\title{Nonlinear Gaussian process tomography with imposed non-negativity constraints on physical quantities for plasma diagnostics.}

\author{Kenji Ueda$^{1,a)}$, Masaki Nishiura$^{1,2.b)}$}
\address{$^{1}$National Institute for Fusion Science, 322-6 Oroshi, Toki, 509-5292 Gifu, Japan\\
$^{2}$Graduate School of Frontier Sciences, The University of Tokyo, Chiba 277-8561, Japan}
\ead{$^{a)}$ueda.kenji@nifs.ac.jp, $^{b)}$nishiura@nifs.ac.jp\\
$^{b)}$Corresponding author}
\vspace{10pt}

\begin{abstract}
We propose a novel tomographic method, nonlinear Gaussian process tomography (nonlinear GPT), that uses the Laplace approximation to impose constraints on non-negative physical quantities, such as the emissivity in plasma optical diagnostics. While positive-valued posteriors have previously been introduced through sampling-based approaches in the original GPT method, our alternative approach implements a logarithmic Gaussian process (log-GP) for faster computation and more natural enforcement of non-negativity.
The effectiveness of the proposed log-GP tomography is demonstrated through a case study using the Ring Trap 1 (RT-1) device, where log-GPT outperforms existing methods, standard GPT, and the Minimum Fisher Information (MFI) methods in terms of reconstruction accuracy. 
The results highlight the effectiveness of nonlinear GPT for imposing physical constraints in applications to an inverse problem.

\end{abstract}

\section{Introduction}
Plasma diagnostics play a crucial role in understanding the internal state and structure of plasmas, as well as in controlling fusion plasmas. 
In-situ diagnostics often provide line-integrated observations, requiring tomographic techniques to obtain cross-sectional distributions for radiation intensity, density, temperature, and velocity.

Inverse problems are generally ill-posed, meaning they often lack a unique solution. 
Therefore, prior information or constraints must be introduced. 
Over the years, numerous tomographic methods have been proposed, with Tikhonov regularization being one of the most prominent. 
The Minimum Fisher Information (MFI) method\cite{L.C.Ingesson_1998,Odstrcil_2016}, which uses Fisher information as the Tikhonov regularization term, is widely utilized in plasma diagnostics\cite{A.Jardin_2021}. 

Meanwhile, Gaussian Process Tomography (GPT) has appeared as a novel approach based on Bayesian inference\cite{J_Svensson_2011}. 
GPT models local physical quantities as a Gaussian process, allowing for the estimation of smooth plasma distributions even from sparse and noisy data. 
This approach has been successfully applied in various contexts, including the estimation of the \(\mathrm{Z_{eff}}\) profile\cite{Kwak2021-yw} and emissivity distributions in soft X-ray measurements\cite{Dong_Li_2013,Wang_T_2018,Wang_T_2018_2}, bolometer diagnostics\cite{Moser2022-zf}, and visible camera diagnostics\cite{Li2021-ce}, as well as electron density profiles in interferometry\cite{Nishizawa2023-mg}.
GPT, as a probabilistic model grounded in Bayesian inference, estimates distributions while quantifying uncertainty through posterior variance. Such frameworks also show promise for the integration of multiple diagnostic datasets\cite{Kwak2021-yw,Wang2019-ac}.

A limitation of existing GPT methods is that they are fundamentally restricted to handling only linear equations and linear constraints. In other words, to derive an analytical solution for the posterior distribution in GPT, the exponents in both the likelihood function and the prior probability must take quadratic forms. To impose the non-negativity constraint, the original GPT study\cite{J_Svensson_2011} proposed aa method of sampling by truncating the negative region from the posterior distribution using Gibbs sampling, which has been applied in several studie\cite{Dong_Li_2021}. 
However, sampling-based methods tend to be computationally expensive, and truncated GPs lack the closed-form characterization inherent in standard GPs, where the mean vector and variance-covariance matrix fully define the statistical properties. As an alternative to Gibbs sampling-based GPT methods, we propose a novel nonlinear Gaussian process tomography (nonlinear GPT) approach using the Laplace approximation.
In particular, we employ a logarithmic Gaussian process (log-GP), which inherently enforces non-negativity.


The outline of this paper is as follows. Section 2 describes conventional GP and GPT. Section 3 introduces log-GP and discusses nonlinear-GPT. Section 4 provides a detailed demonstration of the tomography setup, using the RT-1 experiment as a case study. 
In Section 5, we evaluate the reliability and accuracy of tomography using the nonlinear-GPT method through simulations with phantom data. 
Finally, we discuss in Section 6 and present the conclusion of this study in Section 7.
A previous version of this work was published as a preprint \cite{ueda2024nonlineargaussianprocesstomography}.
\section{General methodology of tomography}

\subsection{Inverse problem}

\begin{figure}[htbp]
    \centering
      \includegraphics[width=60mm]{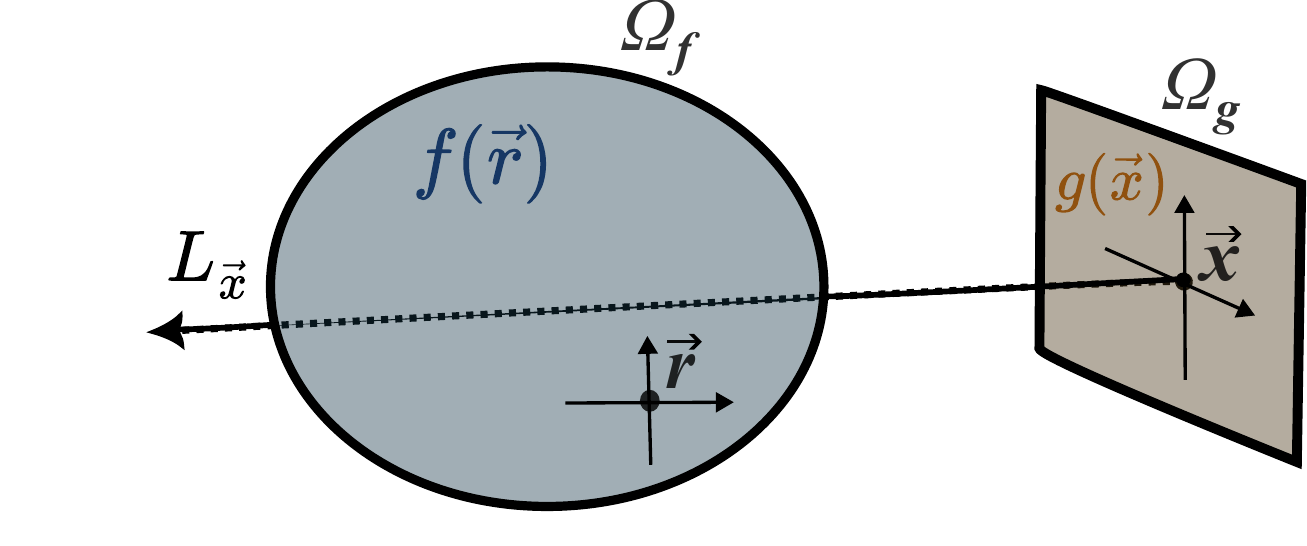}
      \caption{\label{fig: f_g_conceptual.png} Relation between the local quantity \(f\) in the plasma volume \(\Omega_{f}\) and the projected observed quantity \(g\) in a sensor plane \(\Omega_{g}\).}
\end{figure}
As shown in Fig.~\ref{fig: f_g_conceptual.png}, the observed quantity \(g\) is obtained by integrating the local quantity \(f\) along the line of sight, which is represented as:
\begin{eqnarray}\label{eq: f_g_relation}
 g(\vec{x}) = \int_{L_{\vec{x}}} f(\vec{r})\diff l + \epsilon,
\end{eqnarray}
where, \(\vec{x}\) denotes the coordinates in the sensor, \(\vec{r}\) denotes the coordinates of the region where plasma behavior occurs, and \(L_{\vec{x}}\) represents the path of the line of sight (LOS) specified by \(\vec{x}\).
By discretizing Eq.~\ref{eq: f_g_relation}, the relation between the \(N\)-dimensional vector of the local variable \(\bm{f}\) and the \(M\)-dimensional observed data \(\bm{g}^\mathrm{obs}\) is given by adding an observation error term \(\bm{\epsilon}\) as:
\begin{eqnarray}\label{eq: f_g_relation_disc}
    \bm{g}^\mathrm{obs} := \bm{g}(\bm{f}) + \bm{\epsilon} = H \cdot \bm{f} + \bm{\epsilon},
\end{eqnarray}
where, \(H\) is the geometry matrix with dimensions \(M \times N\).
It can also be represented as \(g_{i} = \sum_{j=1}^{N} H_{ij} f_{j}\) for each \(i\).

In the general Tikhonov regularization method, tomography is performed by minimizing the following functional:
\begin{eqnarray}\label{eq: tikonov method}
    \Lambda[\bm{f}] = (\bm{g}^\mathrm{obs} - \bm{g}(\bm{f}))^\mathrm{T} \Sigma_{g}^{-1} (\bm{g}^\mathrm{obs} - \bm{g}(\bm{f})) + \lambda \Lambda_\mathrm{r}[\bm{f}],
\end{eqnarray}
where, \(\Sigma_{g}\) is the covariance matrix of the error term, \(\Lambda_\mathrm{r}[\bm{f}]\) is the regularization term, \(\lambda\) is a regularization parameter, and the superscript \(^\mathrm{T}\) denotes the transpose of the vector or matrix. The MFI method, which uses Fisher information as a regularization term, is often employed in plasma diagnostics. 
Formally, the regularization term is expressed as:
\begin{eqnarray}\label{eq: MFI}
    \Lambda^\mathrm{MFI}_\mathrm{r}[\bm{f}] = \sum_{i=1}^{N} \frac{1}{f_i} (\bm{\nabla} f)_{i} \cdot (\bm{\nabla} f)_{i}. 
\end{eqnarray}
MFI achieves a non-negative solution, since the smaller the amplitude, the stronger the smoothness constraint. However, it requires iterative computation because its regularization term for \(\bm{f}\) is nonlinear.

\subsection{Gaussian process}

When a function \( f \) is modeled as a Gaussian process, we denote this as \( f \sim \mathcal{GP}(\mu, k) \), where \( \mu \) is the mean function and \( k \) is the kernel (or covariance) function. 
For any finite set of input points \( \bm{r} := \{ r_i \}_{i=1}^{N} \), the corresponding function values \( \bm{f} = f(\bm{r}) \) follow a multivariate normal distribution, denoted as \( \bm{f} \sim \mathcal{N}(\mathbf{\mu}, K) \). Here, \( \mathbf{\mu} \) is the mean vector with components \( \mu_i = \mu(r_i) \), and \( K \) is the covariance matrix with entries \( K_{ij} = k(r_i, r_j) \). 
The probability density function(PDF) of the \( N \)-variate normal distribution is given by:
\begin{eqnarray}
    p(\bm{f}) = \mathcal{N}(\bm{f} \mid \bm{\mu}, \Sigma)
    := (2\pi)^{-N/2} |\Sigma|^{-1/2} \exp \left( -\frac{1}{2} (\bm{f} - \bm{\mu})^\mathrm{T} \Sigma^{-1} (\bm{f} - \bm{\mu}) \right).
\end{eqnarray}

The squared exponential (SE) kernel is the \emph{de facto} standard kernel in GP and has the following form:
\begin{eqnarray}
  k_\mathrm{SE}(r, r') = \sigma_{f}^2 \exp \left( -\frac{1}{2} \frac{(r - r')^2}{\ell^2} \right), \nonumber
\end{eqnarray}
where, \( \ell \) is the length scale, which controls the smoothness of \( f \) and defines its spatial scale, and \( \sigma_{f} \) is the output scale that controls its amplitude.
When considering inhomogeneous length scales and multidimensional input spaces \( \vec{r} \), a generalized SE kernel called the Gibbs kernel \cite{Gibbs_1997, Dong_Li_2013} has the form:
\begin{eqnarray}\label{eq: Gibbs_kernel}
  k(\vec{r}, \vec{r}')
  = \sigma_{f}^2  \frac{|\Sigma_{\ell\ell}|^{\frac{1}{4}} |\Sigma_{\ell'\ell'}|^{\frac{1}{4}}}{\left| \frac{1}{2} \Sigma_{\ell\ell} + \frac{1}{2} \Sigma_{\ell'\ell'} \right|^{\frac{1}{2}}}
  \exp \left( -(\vec{r} - \vec{r}')^\mathrm{T} \left( \Sigma_{\ell\ell} + \Sigma_{\ell'\ell'} \right)^{-1} (\vec{r} - \vec{r}') \right).
\end{eqnarray}
Here, \( \Sigma_{\ell\ell} \) is a positive semi-definite matrix that generalizes the distance metric in the space. 
Under the assumption of an isotropic length scale, it becomes a scalar multiple of the identity matrix:
\begin{eqnarray}\label{eq: isotropic_length_scale}  
    \Sigma_{\ell\ell} = \ell^2(\vec{r}) I=
    \left[\begin{array}{cc}
        \ell^2(\vec{r}) & 0 \\
        0 & \ell^2(\vec{r}) 
    \end{array}\right],
\end{eqnarray}
where, \( I \) is the identity matrix and \( \ell(\vec{r}) \) is the position-dependent length scale function.

\subsection{Gaussian process tomography}
In GPT\cite{J_Svensson_2011}, the problem described by Eq.~\ref{eq: f_g_relation_disc} is formulated with the following assumptions:
\begin{eqnarray}
  \bm{\epsilon}\sim \mathcal{N}(\bm{0}, \Sigma_{g}),\quad
  \bm{f}\sim \mathcal{N}(\bm{\mu}_{f}^\mathrm{pri}, K_{f}).\nonumber
\end{eqnarray}
This formulation enables the posterior distribution, \( p(\bm{f} \mid \bm{g}^\mathrm{obs}) \), to be derived using Bayes' rule. 
The posterior distribution is denoted as \(\mathcal{N}(\bm{f} \mid \bm{\mu}_{f}^\mathrm{post}, \Sigma_{f}^\mathrm{post})\).
The known formula for GPT is expressed as follows:
\begin{eqnarray}
  \bm{\mu}_{f}^\mathrm{post} = \bm{\mu}_{f}^\mathrm{pri} +
 (H^\mathrm{T} \Sigma_{g}^{-1} H + K_{f}^{-1})^{-1} H^\mathrm{T} 
  \Sigma_{g}^{-1} (\bm{g}^\mathrm{obs} - H \bm{\mu}_{f}^\mathrm{pri}) 
 \label{eq: mu_for_GPT} \\
  \Sigma_{f}^\mathrm{post} =
 (H^\mathrm{T} \Sigma_{g}^{-1} H + K_{f}^{-1})^{-1}.\nonumber
\end{eqnarray}
In GPT, the term \(\bm{f}^\mathrm{T} K_{f}^{-1} \bm{f} + \log{|K_{f}|}\) serves as the regularization term \(\Lambda_r[\bm{f}]\) in the Tikhonov method, as expressed in Eq.~\ref{eq: tikonov method}. 
This term acts as a prior on the smoothness of \(\bm{f}\), enforcing the expected spatial correlation defined by the kernel \(K_f\).
The advantages of GPT are as follows. 
First, the kernel matrix \(K_f\) allows it to represent a wide range of function shapes and correlations, which is challenging for conventional regularization methods. 
Second, GPT provides a probabilistic framework where the posterior variance matrix \(\Sigma_{f}^\mathrm{post}\) offers a quantification of the uncertainty in the reconstructed solution, enabling error prediction and confidence assessment. 
Third, by applying Bayes’ Occam’s razor, GPT infers the distributions of hyperparameters (e.g., length scales), allowing for both optimal hyperparameter selection and marginalization over these parameters.

\section{Nonlinear Gaussian process tomography}\label{sec: Nonlinear Gaussian Process tomography}   

\subsection{Log-gaussian process}
The accuracy of tomography could be improved by imposing a non-negativity constraint on physical quantities such as plasma radiation, temperature, and densities. 
Methods such as GPT with Gibbs sampling\cite{J_Svensson_2011}, MFI and Maximum Entropy \cite{Ertl1996-hy} can enforce such constraints. 
To address this issue without sampling-based approach, we introduce a latent variable \(\hat{f}\), such that \(f = \exp(\hat{f})\). 
Rather than modeling the physical quantity \(f\) directly as a Gaussian process, we model \(\hat{f}\) as a Gaussian process:
\begin{eqnarray}\label{eq: def_of_logGP}
f = \exp{\hat{f}},\quad \hat{f} \sim \mathcal{GP} 
\end{eqnarray}
In this case, \(f\) is referred to as a "log-Gaussian process" (log-GP) extending the log-normal distribution to the function space and is denoted as \(f \sim \mathrm{log}\mathcal{GP}\).
This formulation often appears as an intermediate variable to enforce non-negativity, for example, in Gaussian process classification \cite{GP_for_ML} and log-Gaussian Cox processes \cite{Moller1998-gd}.

The log-GP has the following key properties:

\begin{itemize}

\item \textbf{Non-negativity}: The variable \(f\) always takes positive values, which can be formally expressed as:
\begin{eqnarray}
    \mathrm{supp}[\;p(f)\;]=(0,+\infty),\quad \mathrm{where} \; f\sim \mathrm{log}\mathcal{GP}.
\end{eqnarray}
Here, \(\mathrm{supp}[\cdot]\) denotes the support of a function, and \(p(f)\) represents PDF of the log-GP. 
The mean function \(\mu\) primarily determines the scale of \(f\), while the output scale \(\sigma_{f}\) controls its dynamic range.
\item \textbf{Closure under multiplication}: The log-GP is closed under multiplication. In contrast to Gaussian processes, which are closed under addition and subtraction, the product or quotient of two log-GPs yields another log-GP:
\begin{eqnarray}
f_1, f_2 \sim \log\mathcal{GP} \Rightarrow f_1 f_2,\, f_1/f_2\sim \log\mathcal{GP}.
\end{eqnarray}
This property is particularly useful in plasma physics, for example, when estimating plasma pressure from density and temperature (\(p = nT\)) or calculating the ratio of spectral intensities (\(I_A / I_B\)).

\item \textbf{Gradient length}: When \(\hat{f}=\log{f}\) has an SE kernel, the variance of \(\frac{1}{f}\frac{df}{dx}\) is the inverse of the length scale \(\ell(x)^{2}\), which is described generally as:
\begin{eqnarray}
    \mathrm{Var}\left(\frac{\bm{\nabla} f}{f}\right) = \mathrm{Var}\left(\bm{\nabla}\hat{f}\right) = \Sigma_{\ell \ell}^{-1},\label{eq: def_of_spartial_length_of_logGP}
\end{eqnarray}
where, \(\mathrm{Var}(\cdot)\) denotes the variance of a variable. In the context of transport analysis\cite{PhysRevLett.86.2325}, \(f/\nabla f\) is defined as the "gradient length," which implies that the standard deviation of the gradient length corresponds to the length scale in log-GP with SE kernels.

\item \textbf{Mean and mode}: The mean value of a log-GP does not equal \(\exp(\mu)\). Similar to a log-normal distribution, the mean is \(\exp(\mu + \frac{1}{2}\sigma_f^2)\), while the mode of the log-GP remains \(\exp(\mu)\).

\item \textbf{Connection to stochastic processes}: The log-GP is associated with geometric Brownian motion and can be obtained as a solution to a stochastic differential equation when spatial directions are correlated based on a kernel function.
\end{itemize}

From the above, log-GP is often more suitable than existing Gaussian processes for estimating physical quantities expressed on a logarithmic scale, such as density, energy, temperature, and radiance. 
In particular, "gradient length" and "closed under multiplication" are advantages that existing GPT\cite{J_Svensson_2011} do not have, which is one of the motivations of our work.
However, even for such physical quantities, the advantage of log-GP diminishes when the relative change is sufficiently small compared to the absolute amount (\(\Delta f \ll f\)).
Furthermore, log-GP cannot be applied to physical quantities taking both positive and negative values, such as velocity.

\subsection{Log-Gaussian Process Tomography}\label{subsec:log-GPT}
When the local physical quantity is modeled as a log-GP,  the relation between the observed data and the local variable can be expressed as: 
\begin{eqnarray}
  \bm{g}^\mathrm{obs} := \bm{g}{(\hat{\bm{f}})} +\bm{\epsilon} =  H \cdot \exp{\hat{\bm{f}}}+\bm{\epsilon}.\nonumber
\end{eqnarray}
Under the assumptions that \(\bm{\epsilon}\sim \mathcal{N}(\bm{0}, \Sigma_{g})\) and \(\bm{\hat{f}}\sim \mathcal{N}(\bm{\mu}^\mathrm{pri}, K_f)\),
the posterior probability of the variable \(\hat{\bm{f}}\) is formulated as:
\begin{eqnarray} \label{eq: logP_log-GPT}
  \underbrace{\log{p(\hat{\bm{f}}\vert \bm{g}^\mathrm{obs})}}_{\mathrm{posterior}}
  &=&\underbrace{-\frac{1}{2}(  H \cdot \exp{\hat{\bm{f}}}-\bm{g}^\mathrm{obs})^\mathrm{T} 
  \Sigma_{g}^{-1}
    (H \cdot \exp{\hat{\bm{f}}}-\bm{g}^\mathrm{obs})}_{\mathrm{likelihood}}\nonumber\\
  & &\underbrace{-\frac{1}{2}(\bm{\hat{f}}-\bm{\mu}^\mathrm{pri})^\mathrm{T} 
  K_{f}^{-1} (\bm{\hat{f}}-\bm{\mu}^\mathrm{pri})}_{\mathrm{prior}} + \,C,\nonumber
\end{eqnarray}
where, \(C\) is a constant term that does not depend on \(\hat{\bm{f}}\).

In the non-linear case, the likelihood function \( p(\bm{g}^{\mathrm{obs}} \mid \bm{f}) \) follows a non-Gaussian distribution, making an analytical solution for the posterior probability intractable. 
As an alternative to obtaining an exact solution, sampling the posterior probability using Markov chain Monte Carlo methods has been proposed\cite{Bielecki2018-mg}.
However, due to the "curse of dimensionality," such methods often require significant computation time, especially in cases where \(N \gg 1000\).
In these situations, an approximate solution can be derived by employing a quasi-probability distribution, such as \( q(\bm{f}) := \mathcal{N}(\bm{f}; \tilde{\bm{\mu}}, \tilde{\Sigma}) \), approximating the true posterior probability \( p(\hat{\bm{f}} \mid \bm{g}^{\mathrm{obs}}) \) as:
\begin{eqnarray}
p(\hat{\bm{f}} \mid \bm{g}^{\mathrm{obs}}) \simeq q(\hat{\bm{f}}) = \mathcal{N}(\hat{\bm{f}} \mid \tilde{\bm{\mu}}, \tilde{\Sigma}),
\end{eqnarray}
with \(\tilde{\bm{\mu}}\) and \(\tilde{\Sigma}\) estimated accordingly. 
Various methods have been proposed for determining how to approximate the posterior \cite{Malte_Kuss_and_Rasmussen_2005}, with known approaches including Laplace approximation \cite{Williams_and_Barbe_1998}, expectation propagation \cite{Minka_2001}, and variational inference based on the  Kullback-Leibler  divergence \cite{Gibbs_and_Mackay_2000,Fujii2018-aj}. 

\subsection{Laplace approximation method}
The log-GP developed here employs the Laplace approximation method, which offers higher computational efficiency compared to other methods, although it is inferior in approximation accuracy. 
We denote the estimated mean and covariance in the Laplace approximation as $\tilde{\mu}_{\mathrm{LA}}$ and $\tilde{\Sigma}_{\mathrm{LA}}$, respectively.

This method involves performing a Taylor series of the log posterior probability $\log{p(\bm{f}|\bm{g}^{\mathrm{obs}})}$ up to the quadratic term around the maximum a posteriori (MAP) estimate $\tilde{\bm{f}}_{\mathrm{MAP}}$, and matching the coefficients with the mean and variance of $q(\bm{f})$.

The details of the Laplace approximation are as follows. 
We define a function $\Psi(\bm{f})$ as:
\begin{eqnarray}
    \Psi(\bm{f}) :=\log{p(\bm{f},\bm{g}^{\mathrm{obs}})}= \log{p(\bm{f}|\bm{g}^{\mathrm{obs}})} + C.
\end{eqnarray}
The Taylor series of $\Psi(\bm{f})$ around a point $\bm{f}=\bm{b}$ yields:
\begin{eqnarray}
\Psi(\bm{f}) = \Psi(\bm{b}) + (\bm{f}-\bm{b})^{\mathrm{T}} \bm{\nabla}\Psi_{(\bm{b})} +\frac{1}{2}(\bm{f}-\bm{b})^{\mathrm{T}} \nabla^2\Psi_{(\bm{b})} (\bm{f}-\bm{b}) +\cdots + C,\nonumber
\end{eqnarray}
where, $\bm{\nabla}\Psi_{(\bm{b})}$ and $\nabla^2\Psi_{(\bm{b})}$ are defined as follows:
\begin{eqnarray*}
\bm{\nabla}\Psi_{(\bm{b})} := \left. \frac{\partial\Psi(\bm{f})}{\partial \bm{f}} \right|_{\bm{f}=\bm{b}}, 
\quad \nabla^2\Psi_{(\bm{b})}:= \left. \frac{\partial^2\Psi(\bm{f})}{\partial \bm{f}\partial \bm{f}} \right|_{\bm{f}=\bm{b}}.
\end{eqnarray*}
By substituting $\bm{b} = \bm{f}_{\mathrm{MAP}}$ and using $\bm{\nabla}\Psi{(\tilde{\bm{f}}_{\mathrm{MAP}})} = \bm{0}$, we obtain:
\begin{eqnarray}\label{eq: expantion_of_Psif_at_fMAP}
\Psi(\bm{f}) \simeq \frac{1}{2}(\bm{f}-\bm{f}_{\mathrm{MAP}})^{\mathrm{T}} \nabla^2\Psi_{(\bm{f}_{\mathrm{MAP}})} (\bm{f}-\bm{f}_{\mathrm{MAP}}) + C.
\end{eqnarray}
By equating Eq.~\ref{eq: expantion_of_Psif_at_fMAP} with the exponent of $q(\bm{f}) = \mathcal{N}(\bm{f}\mid \tilde{\bm{\mu}}^{\mathrm{LA}}, \tilde{\Sigma}^{\mathrm{LA}})$, the following correspondences are derived:
\begin{eqnarray}
\tilde{\bm{\mu}}^{\mathrm{LA}} = \bm{f}_{\mathrm{MAP}}, \quad
\tilde{\Sigma}^{\mathrm{LA}} = -\nabla^2\Psi_{(\bm{f}_{\mathrm{MAP}})}^{-1}.
\end{eqnarray}

\subsection{Solving with Newton-Raphson}\label{subsec:implementation_log-GPT}
To find $\bm{f}_{\mathrm{MAP}}$, one can solve the optimization problem of minimizing $\Psi$. 
For this problem, the most efficient approach is to use the Newton-Raphson method\cite{GP_for_ML}, which is expressed as:
\begin{eqnarray}\label{eq: Newtonmethod}
\tilde{\bm{f}}^{\mathrm{old}} \rightarrow \tilde{\bm{f}}^{\mathrm{new}}; \quad \tilde{\bm{f}}^{\mathrm{new}} = \tilde{\bm{f}}^{\mathrm{old}} 
- \nabla^2\Psi_{(\tilde{\bm{f}}^{\mathrm{old}})}^{-1} \cdot \bm{\nabla} \Psi_{(\tilde{\bm{f}}^{\mathrm{old}})}.
\end{eqnarray}
The update of $\tilde{\bm{f}}$ is iteratively computed until $\bm{\nabla}\Psi_{(\tilde{\bm{f}})} \simeq \bm{0}$ is satisfied. 
When adopting the Newton-Raphson method in Eq.~\ref{eq: Newtonmethod}, both the gradient of the posterior \(\bm{\nabla} \Psi_{(\tilde{\bm{f}})}\) 
and the inverse of the Hessian matrix \(\nabla^2\Psi_{(\tilde{\bm{f}})}^{-1}\) 
need to be calculated in each iteration. 
This inverse matrix is also used to determine the estimated posterior variance \(\tilde{\Sigma}_\mathrm{LA}\) after the iterations converge.

In the case of log-GPT, \(\Psi(\tilde{\bm{f}})\) is expressed as:
\begin{eqnarray}\label{eq: log~posterior_of_logGPT}
    \Psi(\tilde{\bm{f}})=-\frac{1}{2}(\bm{g}{(\tilde{\bm{f}})}-\bm{g}^\mathrm{obs})^\mathrm{T} \Sigma_{g}^{-1}(\bm{g}{(\tilde{\bm{f}})}-\bm{g}^\mathrm{obs})
-\frac{1}{2}(\tilde{\bm{f}}-\bm{\mu}^\mathrm{pri})^\mathrm{T}K^{-1} (\tilde{\bm{f}}-\bm{\mu}^\mathrm{pri})+C.\nonumber\\
\end{eqnarray}
Consequently, \(\bm{\nabla} \Psi_{(\tilde{\bm{f}})}\) and \(\nabla^2 \Psi_{(\tilde{\bm{f}})}\) can be derived for each \(
i\) and \(j\) using matrix calculus as follows: 
\begin{eqnarray}
  \{\bm{\nabla} \Psi_{(\tilde{\bm{f}})} \}_i
  &=&\left. {\frac{\partial \Psi(\hat{\bm{f}})} {\partial \hat{f}_i}}\right|_{\hat{\bm{f}}=\tilde{\bm{f}}}\nonumber\\
  &=&-\{H^\mathrm{T} \Sigma_{g}^{-1}\cdot(\bm{g}{(\tilde{\bm{f}})}
   -\bm{g}^\mathrm{obs})\}_i
   \,\exp {\tilde{f}_{i}}\nonumber\\
  & &- \{K^{-1}\cdot(\tilde{\bm{f}}-\bm{\mu}^\mathrm{pri})\}_{i},\label{eq: nabla_psi_e} \\
  \{\nabla^2 \Psi_{(\tilde{\bm{f}})} \}_{ij}
  &=& \left.\frac{\partial^2 \Psi(\hat{\bm{f}}) }{\partial \hat{f}_{i}\partial \hat{f}_{j}}\right|_{\hat{\bm{f}}=\tilde{\bm{f}}}\nonumber\\
  &=& -\{H^\mathrm{T}  \Sigma_{g}^{-1}H\}_{ij}
  \,\exp {\tilde{f}_{i}}\, \exp {\tilde{f}_{j}} \nonumber\\
   & &-\delta_{ij} \{H^\mathrm{T}  \Sigma_{g}^{-1}\cdot(\bm{g}{(\tilde{\bm{f}})}-\bm{g}^\mathrm{obs})\}_i
   \exp {\tilde{f}_{i}}\nonumber\\
  & &-\{K^{-1} \}_{ij}, \label{eq: laplacian_psi_e}
\end{eqnarray}
where \(\delta_{ij}\) is the Kronecker delta.
The following pseudocode is provided:
\begin{algorithm}
    \caption{Newton-Raphson Laplace Approximation}
    \label{alg1}
    \begin{algorithmic}[1]
    \ENSURE \(\bm{f}^\mathrm{post}  \sim \mathcal{N}(\tilde{\bm{\mu}}^\mathrm{LA} ,\tilde{\Sigma}^\mathrm{LA}) \)
    \STATE \(\tilde{\bm{f}} \leftarrow \tilde{\bm{f}}^\mathrm{\,init}\)
    \WHILE{$\|\bm{\nabla}\Psi \| \neq \bm{0}$}
    \STATE \(\bm{\nabla} \Psi \leftarrow \bm{\nabla}\Psi(\tilde{\bm{f}})\quad \)   \# Eq.~\ref{eq: nabla_psi_e} 
    \STATE \(\nabla^2 \Psi \leftarrow \nabla^2 \Psi(\tilde{\bm{f}})\quad \) \# Eq.~\ref{eq: laplacian_psi_e}  
    \STATE \(\bm{\Delta} \tilde{f} \leftarrow {(\nabla^2\Psi)}^{-1} \cdot \bm{\nabla}\Psi\)
    \STATE \(\tilde{\bm{f}} \leftarrow \tilde{\bm{f}}+\alpha \bm{\Delta} \tilde{f};\quad 0 < \alpha \leq 1\)
    \ENDWHILE
    \STATE \(\tilde{\bm{\mu}}^\mathrm{LA}\leftarrow \tilde{\bm{f}},\quad\tilde{\Sigma}^\mathrm{LA}\leftarrow -({\nabla^2} \Psi )^{-1}\)
    \end{algorithmic}
\end{algorithm}

In Algorithm \ref{alg1}, \(\alpha\) is the step size for the line search. 
Choosing an \(\alpha\) that minimizes \(\|\bm{\nabla} \Psi(\tilde{\bm{f}}+\alpha \bm{\Delta} \tilde{f}) \|\) may accelerates convergence.
The speed of convergence also depends on the initial value of \(\tilde{\bm{f}}\); empirically, it is often faster to assign a scalar value slightly larger than \(\mu\).
Additionally, since exponential calculations are involved during iterations, there is a risk of arithmetic overflow. One way to mitigate this is by setting a maximum value for the absolute change in \(\bm{\Delta}\tilde{f}\).

\subsection{Bayesian Occam’s razor for hyperparameter optimization}\label{subsec:hyperparam}

When we estimate the posterior \(p(\bm{f}\mid\bm{g}^\mathrm{obs})\) in GPT, the hyperparameters \(\bm{\theta}\) that parameterize the prior (\(K_f\)) and the likelihood (\(\Sigma_g\))( such as length scale \(\ell\) and noise scale \(\sigma_g\)) must either be optimized or marginalized over using the evidence \(p(\bm{g}^\mathrm{obs}\mid\bm{\theta})\).
Under the assumption of a flat prior \(p(\bm{\theta})\), evidence is proportional to \(p(\bm{\theta}\mid\bm{g}^\mathrm{obs})\), derived by marginalizing over \(\bm{f}\) as follows:
\begin{eqnarray}\label{eq:evidence}
\mathcal{L}(\bm{\theta}) := p(\bm{g}^{\mathrm{obs}}\mid \bm{\theta}) = \int p(\bm{g}^{\mathrm{obs}}, \bm{f}\mid \bm{\theta}) \mathrm{d} \bm{f}.
\end{eqnarray}  
In our work, because the dimension $M$ of the observed data \(\bm{g}^\mathrm{obs}\) is large, $p(\bm{g}^\mathrm{obs}\mid \bm{\theta})$ becomes sharply peak around optimal point \(\bm{\theta}^{*}\). 
Hence, the integral over \(\bm{\theta}\) can be approximately by the value at \(\bm{\theta}^{*}\) \cite{Wang_T_2018}, which is described as:
\begin{eqnarray}
p(\bm{f}\mid \bm{g}) =  \int p(\bm{f}\mid \bm{g}^\mathrm{obs},\bm{\theta})p( \bm{\theta}\mid\bm{g}^\mathrm{obs} ) \diff \bm{\theta}\simeq   p(\bm{f}\mid \bm{g}^\mathrm{obs},\bm{\theta}^{*})
\end{eqnarray}

For nonlinear-GPT,it is nearly impossible to obtain Eq.~\ref{eq:evidence} analytically. 
However, under the Laplace approximation described as:
\begin{eqnarray*}
    \Psi(\bm{f}) \simeq -\frac{1}{2}(\bm{f}-\tilde{\mu}_{\mathrm{LA}})^\mathrm{T} \tilde{\Sigma}_{\mathrm{LA}}^{-1}(\bm{f}-\tilde{\mu}_{\mathrm{LA}}) + \Psi(\tilde{\mu}_{\mathrm{LA}}),
\end{eqnarray*}
an approximate solution of the evidnce \(\mathcal{L}(\bm{\theta})\) in Eq.~\ref{eq:evidence} can be derived as:
\begin{eqnarray}\label{eq:evidence_lap}
\log\mathcal{L}(\bm{\theta}) \simeq \Psi(\tilde{\mu}_{\mathrm{LA}}) + \frac{1}{2}\log | \tilde{\Sigma}_{\mathrm{LA}}|.
\end{eqnarray}
By substituting Eq.~\ref{eq: log~posterior_of_logGPT} into Eq.~\ref{eq:evidence_lap}, the evidence for log-GPT model is written as:
\begin{eqnarray}
2 \log\mathcal{L}(\bm{\theta}) &\simeq&
-(\bm{g}{(\tilde{\bm{\mu}}^\mathrm{LA})}-\bm{g}^{\mathrm{obs}})^{\mathrm{T}} \Sigma_{g}^{-1} (\bm{g}{(\tilde{\bm{\mu}}^\mathrm{LA})}-\bm{g}^{\mathrm{obs}}) \nonumber\\
& &-(\bm{\tilde{\mu}}^\mathrm{LA}-\bm{\mu}^{\mathrm{pri}})^{\mathrm{T}} K_{f}^{-1} (\bm{\tilde{\mu}}^\mathrm{LA}-\bm{\mu}^{\mathrm{pri}}) \nonumber\\
& &-\log | K |-  \log |\Sigma_{g}| +  \log |\tilde{\Sigma}_{\mathrm{LA}}| - M\log(2\pi). \label{eq: evidence_for_emission}
\end{eqnarray}



\section{Tomography for optical diagnostic at the RT-1 device}

\begin{figure}[ht]
    \centering
    \includegraphics{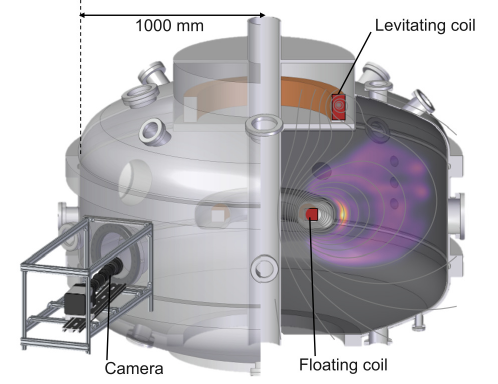}
    \caption{\label{fig: RT-1_camera_conceptual} The magnetospheric plasma device RT-1 and camera system. The camera is set to view a poloidal cross-section of RT-1 torus plasmas and to diagnose an emission from helium ions\cite{K_Ueda_2021}. The emissivity profile from the plasma is overlaid on the magnetic field lines.
    }  
\end{figure}
In this section, we discuss the implementation of log-GPT.
To perform tomography, several factors must be determined: the instrument structure, the instrument's field of view, boundary conditions, the likelihood function, and the geometry matrix \(H\). 
We focus on spectroscopic measurements in the RT-1 device at the University of Tokyo, which is capable of generating laboratory magnetospheric plasmas (Fig.~\ref{fig: RT-1_camera_conceptual}). 
The RT-1 device is composed of a superconducting coil in a vacuum vessel.
The coil is magnetically levitated by an external normal-conducting coil to generate a dipole magnetic field in which the plasma is confined\cite{Z_Yoshida_2006}. 
Using RT-1, We have generated plasmas and observed physical phenomena unique to dipole plasmas, such as plasma pressures exceeding the magnetic field pressure \((\beta > 1)\)\cite{Nishiura_2015}, Van Allen belt-like band structures with high-energy electrons\cite{Nishiura2019-wo}, toroidal flows accelerated by ion heating\cite{Nishiura_2017}, and chorus emissions\cite{Saitoh2024-vh}. 
Due to the absence of magnetic ripple, the RT-1 plasma is purely axisymmetric, which allows for the reconstruction of toroidal cross-sections even from a single image\cite{Naoki_KENMOCHI2019_pfr}.

\subsection{Length scale distribution}
To apply the Gibbs kernel given by Eqs~\ref{eq: Gibbs_kernel}~\ref{eq: isotropic_length_scale} in log-GPT, the length scale function \(\ell(\vec{r}) \) is defined as :
\begin{eqnarray}\label{eq: def of hyperparam}
\ell(\vec{r}) = \hat{\ell}_{s} \ell^{\prime}(\vec{r}),
\end{eqnarray}
where, \(\hat{\ell}_{s}\) is factor of the length scale, and \(\ell^{\prime}(\vec{r})\) represent spatial dependencies

\begin{figure}[ht]
    \centering\textbf{}
    \includegraphics[width=55mm]{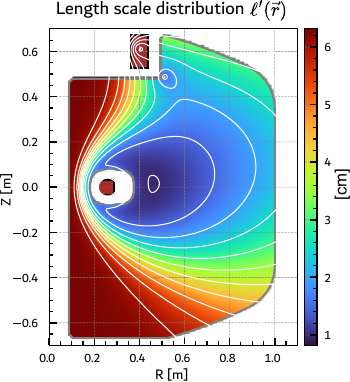}
    \caption{\label{fig: lengthscale_distribution_inRT1} The referenced length scale distribution \(\ell'(\vec{r})\) on the cross-section of RT-1, corresponds to in Eq.~\ref{eq: def of hyperparam}} 
\end{figure}

In the RT-1 device, \(\ell'(\vec{r})\), as shown in Fig.~\ref{fig: lengthscale_distribution_inRT1}, is in the range from 1 cm to 6 cm, which were given empirically from the density and temperature profiles of the RT-1 plasma;  \(\ell'(\vec{r})\) inside the plasma confined area is gradually scaled and connected to the outside of the last closed flux surface (LCFS) and the surface of the superconducting coil case where the plasma lost. 
The length scale factor \(\hat{\ell}_{S}\) is optimized by maximizing the evidence given by Eq.~\ref{eq: evidence_for_emission}.
The use of this inhomogeneous length scale contributes to improved reconstruction accuracy and reduces the size of the vector \(\bm{f}\), as described in the following subsection.

\subsection{Inducing points}
\begin{figure}[ht]
    \centering
    \includegraphics{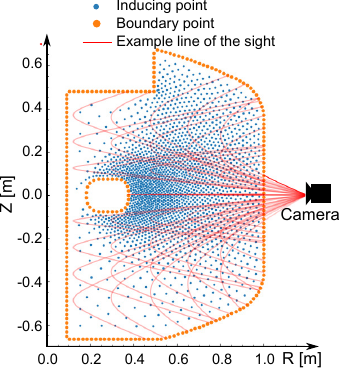}
    \caption{\label{fig: f_scatter_ray} 
    The distribution of inducing points \(\vec{r}^\mathrm{\,idc}\) (the blue points), boundary points \(\vec{r}^\mathrm{\,bd}\) (the orange points), and rays for the  LOS (the red lines) that start from the camera. 
    These rays are projected onto the poloidal cross-section of the RT-1 device. 
    The number of inducing points is 2041, and the number of boundary points is 229 in the setup. The number of rays is reduced for simplicity.
    }
\end{figure}
Functions of physical quantities that follow a GP or log-GP are discretized for numerical computation, which requires defining a set of input spatial points in \(\Omega_{f}\), represented by \(\vec{\bm{r}}\). 
Typically, \(\vec{\bm{r}}\) is arranged on a grid, simplifying the computation of gradients as seen in Eq.~\ref{eq: MFI} and other operations. 
However, in this study, we employed a scattered distribution that is not constrained by a grid structure, as shown in Fig.~\ref{fig: f_scatter_ray}.
In the context of Sparse Gaussian Process Regression\cite{Snelson_Edward_2005}, inducing points are defined as \(\bm{f}^\mathrm{idc} = f(\vec{\bm{r}}^\mathrm{\,idc})\), where \(\vec{\bm{r}}^\mathrm{\,idc}\) is a set of inducing input points. 
The inducing points are placed so that the distance between adjacent points is approximately equal to the referenced length scale \(\ell'(\vec{r})\). 
This is based on the principle that the minimum interval of input point, \(\Delta x\), required to maintain the accuracy of interpolation in the SE kernel is nearly equal to the length scale \(\ell\). 
In other words, the distribution of the inducing points determines the lower limit of the length scale. 
Thus, while searching for the optimal hyperparameters, the length scale factor \(\ell_{S}\) (defined in \ref{eq: def of hyperparam}) must be greater than one. 
Such usage of arbitrary inducing points is expected to significantly reduce the size of the local variable \(\bm{f}\).
This advantage is significant for nonlinear GPT, as the iterative process often requires substantial computational time.

\subsection{Geometry matrix}
The variable \(\bm{f}^\mathrm{idc}\) induces an approximate value at an arbitrary point through the inner product with the weighting functions, as indicated by the equation: \(\tilde{f}(\vec{r})\simeq  \hat{\bm{w}}(\vec{r})^\mathrm{T} \cdot \bm{f}^\mathrm{idc}\). 
Substituting this variable into Eq.~\ref{eq: f_g_relation} yields the following relationship between \(\bm{g}\) and \(\bm{f}^\mathrm{idc}\):
\begin{eqnarray}
    g(\vec{x}_i) \simeq \int_{L_i} \sum_{j=1}^{N} \hat{\omega}_j(\vec{r}) {f}^\mathrm{idc}_j \diff l =
    \sum_{j=1}^{N} \left( \int_{L_i} \hat{\omega}_j(\vec{r})\diff l \right)  {f}^\mathrm{idc}_j. 
\end{eqnarray}
Considering the equation \(\bm{g} = H\cdot \bm{f}^\mathrm{idc}\), the matrix \(H\) is derived as:
\begin{eqnarray}
    H_{ij} \simeq \int_{L_i} \hat{\omega}_j(\vec{r})\diff l,
\end{eqnarray}
where, \(i\) denotes the index of the LOS corresponding to the pixel on the sensor plane, and \(j\) denotes the index of the inducing points. 
There are several approaches to define the weighting function. 
For instance, the kernel interpolation method, which is expressed as \(\hat{{\omega}}_{j}(\vec{r}) = \sum_{i}^{N}k(\vec{r},\vec{r}^\mathrm{\,idc}_{i})  \{K^\mathrm{idc}\}^{-1}_{ij}\), might be consistent in terms of the function space\cite{Xu2024-iu}.  
Alternatively, when the dimension of the observed data \(M\) is very large, as in the case of imaging measurements such as our research, we employed the ratio of the kernel functions as the weighting function to ensure the sparsity of the matrix. The weighting functions are defined as:
\begin{eqnarray}
    \hat{\omega}_j(\vec{r}) = \frac{k(\vec{r},\vec{r}^\mathrm{\,idc}_j)}{\sum_{k=1}^{N} k(\vec{r},\vec{r}^\mathrm{\,idc}_{k})},
\end{eqnarray}
where, \(k(\vec{r},\vec{r})\) is the generalized Gibbs kernel function defined in Eq.~\ref{eq: Gibbs_kernel}, and \(\ell(\vec{r})\) is set to be the same distribution as the length scale in Fig.~\ref{fig: lengthscale_distribution_inRT1}.
The function \(k(\vec{r},\vec{r}^\mathrm{\,idc})\) rapidly approaches zero when the distance between \(\vec{r}^\mathrm{\,idc}\) and \(\vec{r}\) exceeds several times of \({\ell}'(\vec{r})\). 
Consequently, the weighting function \(\hat{\omega}_j(\vec{r})\) becomes nearly zero, which results in sparsity in the geometry matrix.

\subsection{Boundary conditions}
The boundary conditions are imposed such that the value of the local quantities is zero at the container wall. 
One way to incorporate boundary conditions within Bayesian estimation is to construct conditional probabilities based on prior probabilities.

In this approach, the joint probability distribution of \(f(\vec{\bm{r}}^\mathrm{\,idc})\) and \(f(\vec{\bm{r}}^\mathrm{\,bd})\) is considered as:
\begin{eqnarray*}
\left[
\begin{array}{c}
\bm{f}^\mathrm{idc} \\
\bm{f}^\mathrm{bd}
\end{array}
\right]
\sim \mathcal{N}\left(
\left[
\begin{array}{c}
\bm{\mu}^\mathrm{idc} \\
\bm{\mu}^\mathrm{bd}
\end{array}
\right],
\left[
\begin{array}{cc}
K^\mathrm{idc} & K^\mathrm{idc,bd} \\
K^\mathrm{bd,idc} & K^\mathrm{bd}+\sigma^2 I
\end{array}
\right]
\right),
\end{eqnarray*}
and then the conditional probability is calculated given \(f(\vec{\bm{r}}^\mathrm{\,bd}) \sim \mathcal{N}(\bm{y}^\mathrm{\,bd}, \sigma^2 I)\). 
Using the well-known formula for the conditional probability of a multivariate normal distribution, the mean and covariance of \(\bm{f}^\mathrm{idc}\) are updated as:
\begin{eqnarray}
    \bm{\mu}^{\mathrm{idc}\mid\mathrm{bd}} &=& \bm{\mu}^\mathrm{idc} +  K^\mathrm{idc,bd}  (K^\mathrm{bd}+ \sigma^2 I)^{-1}  (\bm{y}^\mathrm{bd} - \bm{\mu}^\mathrm{bd} ),\nonumber\\
    K^{\mathrm{idc}\mid\mathrm{bd}}  &=& K^\mathrm{idc}- K^\mathrm{idc,bd} K^\mathrm{bd}+\sigma^2 I)^{-1} K^\mathrm{bd,idc}.
\end{eqnarray}
Here, \(K^\mathrm{idc,bd}\) and \(K^\mathrm{bd,idc}\) are the covariance matrices between the interior points (inducing points) and the boundary points, defined as \(K^\mathrm{idc,bd}_{ij} = k(\vec{r}^\mathrm{\,idc}_i,\vec{r}^\mathrm{\,bd}_j)\) and \(K^\mathrm{bd,idc} = {(K^\mathrm{idc,bd}})^\mathrm{T}\). \(K^\mathrm{bd}\) and \(K^\mathrm{idc}\) are the covariance matrices within the boundary and interior points, respectively.

It is impossible to make \(f\equiv\exp{\hat{f}}\) completely zero because this would imply that \(\hat{f}\) is negative infinity. 
Therefore, in the log-GPT framework, it is reasonable to set \(\bm{y}^\mathrm{bd}-\bm{\mu}^\mathrm{bd}\) to a value around \(-3\) to \(-5\) times \(\sigma_f\).

Regarding the prior mean values \(\bm{\mu}^\mathrm{pri}\), a simple approach is to normalize the observation data \(\bm{g}^\mathrm{obs}\) such that the average of \(\bm{\mu}^\mathrm{post}\) will be approximately zero, instead of assigning a specific value. In this case, \(\bm{g}^\mathrm{obs} \) may be divided by \(\langle \bm{g}^\mathrm{obs} \rangle / \langle H \rangle\) before performing the tomography, where \(\langle \bm{g}^\mathrm{obs} \rangle\) is given by \(\frac{1}{M}\sum_{i} g^\mathrm{obs}_{i}\) and \(\langle H \rangle\) is given by \( \frac{1}{M}\sum_{i,j} H_{ij}\). 

\subsection{Likelihood Function}

Modeling the likelihood function requires determining the variance of the error, \(\Sigma_{g}\), before performing the tomography. 
There are several approaches to specifying the variance. 
The simplest approach is to use a scalar multiple of the identity matrix\cite{Dong_Li_2013}, which is appropriate when the errors are independent and identically distributed.
A model such as \( \sigma(\vec{x}) \propto g^\mathrm{obs}(\vec{x})\) is used \cite{Wang_T_2018,Wang_T_2018_2} when the error is assumed to be proportional to the intensity of the signal, and a model such as \(\sigma(\vec{x}) \propto \sqrt{g^\mathrm{obs}(\vec{x})}\) is also considered when shot noise is dominant. 
Alternatively, if the error is known to be spatially correlated, a variance-covariance matrix with off-diagonal terms can be used for \(\Sigma_g\). 

In the tests with phantom data described below, since white Gaussian noise was artificially added, we used following model: 
\begin{eqnarray}
    \Sigma_g = \sigma_{S}^2 I,
\end{eqnarray}
where, \(\sigma_{S}\) is one of the hyperparameters to be optimized.

\section{Phantom test}

\subsection{log-GPT with phantom data}
\begin{figure}[htb]
    \centering
    \includegraphics{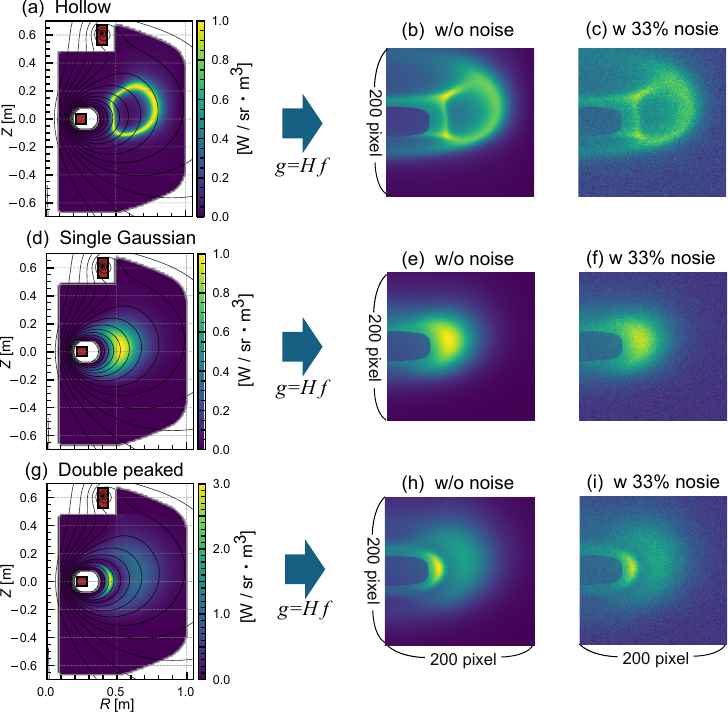}
       \caption{\label{fig: phantom_data}  The phantom distributions and their projected images. The left column, (a), (d), and (g), represent the distributions labeled as ``Hollow,'' ``Single Gaussian,'' and ``Double peaked,'' respectively. The middle column, (b), (e), and (h), shows the projected images obtained using the geometry matrix \(H\) applied to the distributions in the left column. The right column, (c), (f), and (i), displays the results with 33\% noise (for example) added to the images.}
\end{figure}

In this subsection, we evaluate the performance of tomography with log-GPT using synthetic ``phantom'' distributions. 
Figure~\ref{fig: phantom_data} shows three typical distributions we use for validation: (a) Hollow, (d) Single Gaussian, and (g) Double peaked. 
These distributions are chosen to mimic emission patterns commonly observed in torus plasmas, though they are artificially defined here for clarity (see \ref{appendix: Definition of Phantom distribution} for details).
For instance, the Hollow distribution reflects typical edge emission patterns often seen in large toroidal devices, 
while the Single Gaussian distribution is inspired by electron density distributions reported in RT-1~\cite{Nishiura2019-wo}. 
The Double peaked distribution represents visible-light patterns frequently observed in RT-1 experiments~\cite{K_Ueda_2021,Naoki_KENMOCHI2019_pfr,Nakamura_2018}.

It is important to emphasize that our method does not specify a single physical quantity \textit{a priori}. 
Nonetheless, we are particularly interested in reconstructing the emissivity of the 486\,nm line of \(\mathrm{He}^+\) emission in RT-1 for coherence imaging spectroscopy~\cite{K_Ueda_2021,Nakamura_2018}. 
Hence, for illustration, we assume emissivity units of \(\mathrm{W \cdot sr^{-1} \cdot m^{-3}}\) and normalize the scales for simplicity.

The test input data, denoted as \(\bm{g}^\mathrm{inp}\), were generated by projecting the phantom data using the geometry matrix \(H\) with white Gaussian noise at levels of 1\%, 3.3\%, 10\%, 33\%, and 100\%. 
The noise level is defined as the ratio of the standard deviation of the noise \(\sigma_\epsilon\) to the mean value of the projected data, and is calculated as:
\begin{eqnarray}
\mathrm{noise\,level}\;(\%) = 100\times \sigma_{\epsilon}/{\langle \bm{g} \rangle}.
\end{eqnarray}

\begin{figure}[htbp]
    \centering
    \includegraphics{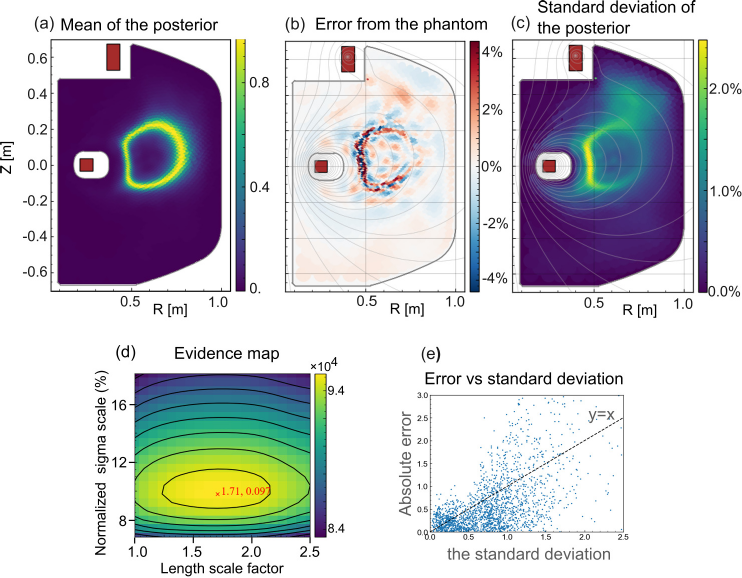} 
    \caption{\label{fig: phantom_result_e} Results of log-Gaussian process tomography when the phantom distribution is the Hollow and the noise level is 10\%. (a) Mean of the posterior probability of the local emissivity.
    (b) Error between the mean posterior value and the true value. (c) Standard deviation of the posterior probability of the local emissivity. 
    (d) Mapping of the evidence, where the optimal point for the hyperparameters has a normalized sigma scale of 9.7\% and a length scale factor of 1.71.  
    (e) Relation between the standard deviation (c) and the error (b).}
\end{figure}

Fig.~\ref{fig: phantom_result_e} shows tomographic results of log-GPT for the Hollow distribution phantom with the 10\% noise level.
The hyperparameters of this model are determined by mapping the evidence given in Eq.~\ref{eq: evidence_for_emission} with respect to the length scale factor \(\ell_{S}\) and the sigma scale \(\sigma_{S}\), as shown in the Fig.~\ref{fig: phantom_result_e}(d). 
The optimized values are \(\hat{\ell}_{S} = 1.71\) and \(\sigma_{S} = 9.7\%\), indicating that the optimized length scale corresponds to 1.71 times the referenced length scale distribution in Fig.~\ref{fig: lengthscale_distribution_inRT1}, and the optimized sigma scale is almost the same as the noise level of 10\%. 
The strong effect of the sigma scale on the evidence may be due to the fact that the number of observation data \(M\) is considerably larger than the number of inducing points \(N\) (40000 vs. 2041), making the first and fourth terms in Eq.~\ref{eq: evidence_for_emission} dominant.
Fig.~\ref{fig: phantom_result_e}(a) shows the mean of the posterior probability of the local emissivity, which is the tomographic result.
Fig.~\ref{fig: phantom_result_e}(b) shows the residual error distribution from the true distribution. Fig.~\ref{fig: phantom_result_e}(c) shows the standard deviation of the posterior. 
The standard deviation reflects the uncertainty in the emissivity distribution, as shown in Fig.~\ref{fig: phantom_result_e}(e), which is important for evaluating the tomographic results.

\subsection{Comparison with other methods}

To evaluate the performance of the proposed log-GP method, we compared it with the standard GPT without Gibbs sampling and the MFI. 
The log-GPTs with uniform length scale and non-uniform length scale were used to investigate the influence of the length scale distribution.
The input data was based on the Hollow, the Single Gaussian, and the Double peaked, as shown in Fig.~\ref{fig: phantom_data}. 
The noise levels range from 1\% to 100\%. 
The standard GPT and the log-GPT are performed with the same setting on the non-uniform length scale distribution in Fig.~\ref{fig: lengthscale_distribution_inRT1} and the inducing points in Fig.~\ref{fig: f_scatter_ray}, although the hyperparameters, \(\hat{\ell}_S\), are selected individually.
The log-GP with uniform length scale and MFI use the same input points with a grid structure, where the interval is 1 cm and the size is \(80 \times 100 = 8000\).
The sigma scales \(\sigma_S\) for each condition of the GPTs were assigned a value corresponding to the input noise levels without optimizing the evidence.
The length scale factor \(\hat{\ell}_S\) for each method was chosen in Table.~\ref{tab: hyperparameter of length scale} as the optimal value for the 10\% noise level.
The regularization parameter \(\lambda\) in MFI was selected to minimize the total reconstruction error in each condition, which is defined as follows:
\begin{eqnarray}
    \mathrm{total\;error}\;(\%)= 100\times \sqrt{\langle|\tilde{f}-{f}_\mathrm{true}|^2\rangle}/\sqrt{\langle{{f}_\mathrm{true}^2}\rangle}.
    \label{eq: def_of_error}
\end{eqnarray}

\begin{figure}[htbp]
\centering
    \includegraphics{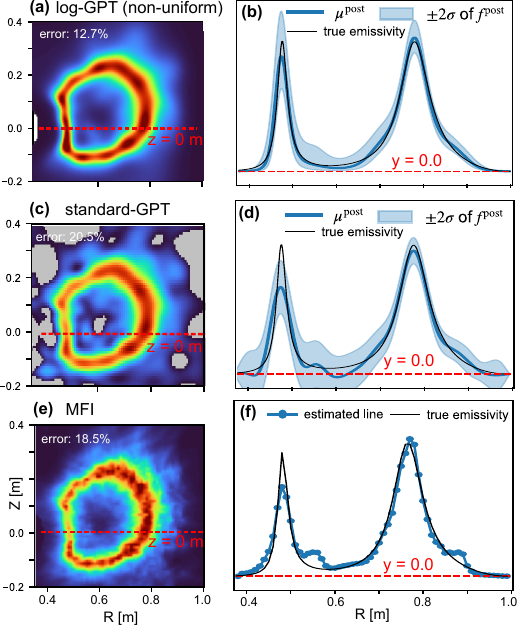}
    \caption{The results of the comparison between three methods, where the noise level of the input image is 100\% in each method. The top (a,b), middle (c,d) and bottom (e,f) rows show the results of log GPT, standard GPT and MFI, respectively.
     The left figures (a, c, e) show the two-dimensional distributions of the estimated values, and the right figures (b, d, f) show the radial profiles of the emissivity at \(Z = 0\) m. The regions below 0 are masked in gray, as shown in (b). The black lines on the right represent the true distribution, and the red dashed lines indicate the zero value.}
    \label{fig: compare_mothod_profile}
\end{figure}

The results of the comparison are summarized in Table~\ref{tab: compareing}, which demonstrates that the proposed method, log-GPT (non-uniform), has the smallest total error in almost all case. 
It is noteworthy that although the MFI deliberately optimizes the hyperparameter for the best accuracy, log-GPT still outperforms it. 
However, it should be noted that comparing GPT to MFI is not entirely fair because the length scale distribution of GPT, \({\ell}(\vec{r})\), is adjusted to fit any local distribution associated with RT-1, as shown in Fig.~\ref{fig: lengthscale_distribution_inRT1}. Indeed, there are some cases where MFI is more accurate than log-GPT (uniform).

Fig.~\ref{fig: compare_mothod_profile} specifically displays the tomographic results at a 100\% noise level for the three methods. 
Comparing Figs.~\ref{fig: compare_mothod_profile}(a), (c), and (e), the standard GPT produces a locally negative solution for both the mean and confidence interval, and the MFI exhibits a noise-like distribution. 
It is evident that the MFI solution would be smoother with a larger \(\lambda\), but this would likely result in a larger error. 
From Fig.~\ref{fig: compare_mothod_profile}(c), we can recognize that the standard GPT solution violates the physical constraint that it is locally greater than zero. 
Even if the mean value is greater than zero, as shown in Fig.~\ref{fig: compare_mothod_profile}(f), the MFI method may still violate the constraint when the variance is considered, which can complicate subsequent analyses, including error propagation.

\begin{table}[htb]  
    \caption{Comparison of accuracy between four tomographic methods: standard GPT, log-GPT with non-uniform length scale, log-GPT with uniform length scale, and MFI.
     The types are separated into the Hollow, the Single Gaussian, and Double peaked, with the noise levels varied from 1\% to 100\%.  
     Each cell shows the mean $\pm$ standard deviation of the total reconstruction error (\%), as defined in Eq.,\ref{eq: def_of_error}, calculated over more than ten random input images. 
     For each GPT, hyperparamer are selected according to \ref{tab: hyperparameter of length scale}. Bold numbers indicate the most accurate value under the same condition.}  
    \label{tab: compareing}
    \centering
    \scriptsize
    \begin{tabular}{c|c|c||ccccc} 
        \toprule
       & &  &\multicolumn{5}{c}{Noise level} \\ 
       Method &  Structure &  Phantom  & 1\% & 3.3\%  & 10\%  & 33\%   & 100\%\\
        \midrule
      Standard GPT    & non-uniform &\multirow{4}{*}{Hollow}& \(3.4\pm0.3\)& \(6.8\pm0.8\)& \(10.7\pm1.1\)& \(16.0\pm0.8\)& \(25.4\pm1.5\)\\ 
      log-GPT    & non-uniform     &  & $\bm{1.5}\pm0.1$ & $\bm{2.2}\pm0.1$ & $\bm{3.5}\pm{0.15}$ & $\bm{6.9}\pm{0.3}$ & $\bm{13.8}\pm{1.0}$ \\  
      log-GPT  & uniform &  & $3.2\pm0.1$ &  $4.1\pm0.2$ & $5.4\pm0.2$  & $8.2\pm0.2$  & $\bm{13.8}\pm0.9$ \\  
      MFI      & uniform &  & $2.1\pm0.1$ & $3.5\pm0.1$&$6.4\pm0.2$& $11.4\pm0.4$ & $18.5\pm0.6$\\ 
    \midrule
      Standard GPT    & non-uniform &\multirow{4}{*}{\begin{tabular}{c}Single \\Gausian\end{tabular}} & $1.7\pm0.2$  &  $4.2\pm0.7$ & $6.7\pm0.8$ & $10.0\pm0.9$ & $15.1\pm0.9$\\ 
      log-GPT    & non-uniform     & & $\bm{1.22}\pm0.03$  &  $\bm{1.75}\pm0.04$ & $\bm{2.8}\pm0.1$ & $\bm{5.5}\pm0.2$ & $11.1\pm0.6$\\  
      log-GPT  & uniform & &  $3.69\pm0.06$  & $4.4\pm0.1$  & $6.0\pm0.1$ &$9.5\pm0.2$   & $16.0\pm0.7$  \\  
      MFI      & uniform &  &$1.8\pm0.02$& $2.8\pm0.04$ & $4.0\pm0.1$ & $6.6\pm0.1$ & $\bm{10.2}\pm0.5$\\ 
    \midrule
      Standard GPT    & non-uniform &\multirow{4}{*}{\begin{tabular}{c}Double \\peaked\end{tabular}}  & $\bm{2.4}\pm0.3$   &  $4.3\pm0.6$ & $7.2\pm0.6$ & $11.2\pm0.7$ & $20.3\pm1.0$\\ 
      log-GPT    & non-uniform     & & $2.9\pm0.1$   & $\bm{4.2}\pm0.1$ & $\bm{5.4}\pm0.2$  & $\bm{8.6}\pm0.3$ & $\bm{16.0}\pm0.7$  \\
      log-GPT  & uniform & & $4.3\pm0.2$ & $6.5\pm0.4$ & $8.3\pm0.4$ & $14.9\pm0.4$ & $26.1\pm0.8$ \\  
      MFI      & uniform &  &$3.68\pm0.05$ &$4.97\pm0.05$&$7.2\pm0.1$& $12.2\pm0.3$ &$21.1\pm0.9$\\ 
          \bottomrule
    \end{tabular}
\end{table}

\section{Discussion}
The accuracy of tomographic methods using Bayesian theorem depends on how well the prior distribution captures the possible plasma distributions. The superior accuracy of the log-GPT observed in this study is likely due to the fact that it inherently represents plasma distributions more naturally than the standard GPT. However, it should be noted that the proposed method requires more computational time compared to conventional methods, mainly due to the iterative nature of the Newton-Raphson method.

The log-GPT and MFI methods share a conceptual similarity; in the MFI method, the regularization weight for smoothing is inversely proportional to the amplitude (Eq.~\ref{eq: MFI}), while in the log-GPT method, the standard deviation of the gradient is proportional to the amplitude (Eq.~\ref{eq: def_of_spartial_length_of_logGP}).
However, GPT allows us to optimize the regularization intensity and smoothness through the sigma scale and length scaleparameters independently, which is advantageous for inverse transform of localized structures with noise signals. 
In this respect, log-GPT combines the advantages of both GPT and MFI, providing a more flexible and robust framework for plasma tomography.

A key strength of the logarithmic transform is that it handles physical quantities constrained to be positive or multiplicative. 
For instance, plasma pressure can be expressed as the product of density and temperature ($p = nT$), while bremsstrahlung radiation scales as \( \propto n_e^2 T_e^{-1/2} Z_\mathrm{eff}^2\exp(-h\nu/T_e)\)\cite{Nishiura2019-wo}. 
Incorporating such multiplicative relationships into the log-domain can naturally improve plasma parameter estimation through log-GPs. 
Additionally, many diagnostic methods, such as optical emission spectroscopy, infer plasma parameters from the ratio of line intensities\cite{SCHWEER1992174,GOTO2003331,Howard_2003}. 
Because a ratio of two log-GP-modeled quantities remains log-GP, it is straightforward to propagate uncertainties and properly evaluate confidence intervals in the derived physical parameters.
Furthermore, the gradient lengths of temperature and density profiles (\(L_T^{-1}=
\nabla\log(T)\) and \(L_n^{-1}=\nabla\log(n)\)) are key parameters in plasma transport analysis\cite{PhysRevLett.86.2325}. 
In the log-GP framework, these gradient lengths can be directly encoded in the kernel hyperparameters (length scales), allowing one to impose physically meaningful priors on spatial variation.

Finally, while the present study just focuses on the tomographic reconstruction of visible light measurements from a tangential view in the RT-1 instrument, it is based on idealized numerical experiments. 
In actual visible light measurements, the effect of reflections from wall should be carefully considered in future studies to ensure the applicability of the method to real-world   scenarios.

\section{Conclusion}

We proposed a novel tomographic method, nonlinear-GPT using Laplace approximation, focusing on implementing log-GPT to ensure non-negativity. 
Inducing points were used to improve computational efficiency, and the local emissivity on the poloidal cross-section was reconstructed from line-integrated emissivity data. 
The effectiveness and performance of this approach were demonstrated through a case study based on the RT-1 device configuration. 
Accuracy validation using phantom data revealed that the proposed method outperforms existing tomographic methods, including standard GPT and MFI.

\section*{Acknowledgments}
This work was supported by JSPS KAKENHI Grant Numbers JP17H01177, JP19KK0073, and JP23K25857. It was also supported by JST SPRING Grant Number. JPMJSP2108.
\clearpage
\section*{References}
\bibliographystyle{iopart-num}  
\bibliography{your-bib-file}  

\appendix

\section{Definition of phantom distributions}
In this section, we describe the definitions of the three phantom distributions used in our study: the Hollow distribution, the Single Gaussian distribution, and the Double peaked distribution. Each distribution is expressed in terms of the magnetic flux function \(\psi(r,z)\) and the magnetic field strength \(B(r,z)\) in RT-1.

The Single Gaussian and Double Peaked distributions are both built using the function \(\mathrm{G}\) introduced in Ref.~\cite{Nishiura2019-wo}, which is parameterized by \(a\), \(b\), and \(r_\mathrm{peak}\):
\begin{eqnarray}\label{appendix: Definition of Phantom distribution}
    \mathrm{G}(r,z;\, a, b, r_\mathrm{peak}) &=&
    \exp\Bigl(-a\,\hat{\psi}(r,z;\,r_\mathrm{peak})^2\Bigr)
    \,\hat{B}(r,z)^{-b},
\end{eqnarray}
where,
\begin{eqnarray}
    \hat{\psi}(r,z;\,r_\mathrm{peak}) 
    &=& \frac{\psi(r,z)-\psi(r_\mathrm{peak},0)}{\psi_0}, \quad
    \hat{B}(r,z) 
    \;=\; \frac{B(r,z)}{B_0(r,z)}.
\end{eqnarray}
Here, \(\hat{\psi}(r,z;\,r_\mathrm{peak})\) is the normalized flux function, centered at \(r_\mathrm{peak}\) and divided by \(\psi_0 = \psi(1\,\mathrm{m},0\,\mathrm{m})\). The term \(\hat{B}(r,z)\) is the local magnetic mirror ratio, where \(B_0(r,z)\) denotes the minimum magnetic field strength on the same flux surface \(\psi(r,z)\). The parameter \(b\) represents the strength of the mirror effect in the dipole magnetic field.

For this phantom test, the Single Gaussian and Double peaked distributions are defined as follows:
\begin{eqnarray}
    f_\mathrm{Single\,Gaussian}(r,z) &=& \mathrm{G}\bigl(r,z;\,a=6.3,\,b=1,\,r_\mathrm{peak}=0.53\,\mathrm{m}\bigr), \\
    f_\mathrm{Double\,Peaked}(r,z) &=& 
    \mathrm{G}\bigl(r,z;\,a=15,\,b=0.65,\,r_\mathrm{peak}=0.65\,\mathrm{m}\bigr) \nonumber\\
    \;&+& 3\,\mathrm{G}\bigl(r,z;\,a=35,\,b=2,\,r_\mathrm{peak}=0.45\,\mathrm{m}\bigr).
\end{eqnarray}

The Hollow distribution is given by  Cauchy distribuion with a ring-like shape:
\begin{eqnarray}
    f_\mathrm{Hollow}(r,z) &=& 
    \frac{0.06^2}{\Bigl(\sqrt{\hat{\psi}\bigl(r,z; 0.58\,\mathrm{m}\bigr)^2 \;+\; \Bigl(1-\frac{0.8}{\hat{B}(r,z)}\Bigr)^2} - 0.42\Bigr)^2 \;+\; 0.06^2}.
\end{eqnarray}

\begin{figure}[ht]
    \centering
    \includegraphics{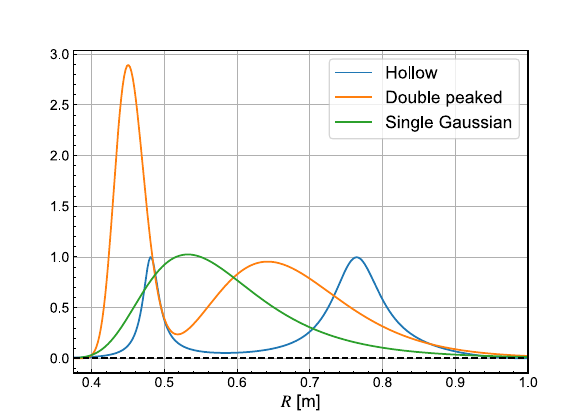}%
    \caption{\label{fig: phantom_1D} Radial profiles of the three phantom distributions on the midplane (\(z = 0\,\mathrm{m}\)).}
\end{figure}

Figure~\ref{fig: phantom_1D} shows the radial profiles of these three distributions on the midplane \((z=0\,\mathrm{m})\).

\newpage
\section{Hyperprameters for the phantom test}
Table~\ref{tab: hyperparameter of length scale} summarizes the length-scale parameters obtained via evidence optimization. In principle, the optimal length scale depends on the noise level of the input images (indeed, it tends to increase slightly with higher noise levels). However, in this test, the length scale is held fixed for all noise levels; we simply used the values optimized under the 10\% noise condition.

\begin{table}[htb]
    \caption{Hyperparameters for the length scale in standard-GPT and log-GPT. Here, $\hat{\ell}_s$ denotes the length-scale factor, while $\ell$ indicates a uniform length scale. Each value was optimized under the 10\% noise condition.}
    \label{tab: hyperparameter of length scale}
    \centering
    \footnotesize
    \begin{tabular}{c||cccc} 
      \toprule
       & \begin{tabular}{c} standard-GPT \\ (non-uniform) \end{tabular} 
       & \begin{tabular}{c} \textbf{log-GPT} \\ (non-uniform) \end{tabular}
       & \begin{tabular}{c} \textbf{log-GPT} \\ (uniform) \end{tabular} \\
      \midrule
       Hollow & \(\hat{\ell}_s = 1.48\) & \(\hat{\ell}_s = 1.71\) & \(\ell = 4.4\,\mathrm{cm}\) \\
       Single Gaussian & \(\hat{\ell}_s = 1.90\) & \(\hat{\ell}_s = 1.84\) & \(\ell = 3.2\,\mathrm{cm}\) \\
       Double Peaked & \(\hat{\ell}_s = 1.70\) & \(\hat{\ell}_s = 1.76\) & \(\ell = 2.7\,\mathrm{cm}\) \\
      \bottomrule
    \end{tabular}
\end{table}
\end{document}